\documentclass[twocolumn,showpacs,preprintnumbers,amsmath,amssymb]{revtex4}

\newcommand{\ds}{\displaystyle}

\newcommand{\be}{\begin{equation}}
\newcommand{\en}{\end{equation}}
\newcommand{\bea}{\begin{eqnarray}}
\newcommand{\ena}{\end{eqnarray}}

\usepackage{graphicx}
\usepackage{dcolumn}
\usepackage{bm}

\begin{document}

\title{Accelerated closed universes in scalar-tensor theories}

\author{ Sergio del Campo\,$^{a}$
  and Patricio Salgado\,$^{b ,\,\, c}$}
\address{$^a$  Instituto de F\'\i sica, Pontificia Universidad Cat\'olica de\\
Valpara\'\i so, Casilla 4059, Valpara\'\i so, Chile.\\
$^b$ Departamento de F\'\i sica,\\
Universidad de Concepci\'on, Casilla 160-C, Concepci\'on, Chile.\\
$^c$ Max-Planck Institut, f\"{u}r Str\"{o}mungsforschung \\
Bunsenstrasse 10, D-37073, G\"{o}ttingen, Germany}

\date{\today}

\begin{abstract}
We describe an accelerating universe model in the context of a
scalar-tensor theory. This model is intrinsically closed, and is
filled with quintessence-like scalar field components, in addition
to the Cold Dark Matter component. With a background geometry
specified by the Friedman-Robertson-Walker metric, we establish
conditions under which this closed cosmological model, described
in a scalar-tensor theory, may look flat in a genuine
Jordan-Brans-Dicke theory. Both models become indistinguishable at
low enough redshift.
\end{abstract}

\maketitle

\section{\label{sec:level1} Introduction}

Recent astronomical observations conclude that the matter density
related to baryonic matter and to nonbaryonic cold dark matter, is
much less than the critical density value~\cite{Wh-etal}. From
this it may be concluded that either the universe is open or that
there are some other matter components which make the total matter
density parameter, $\Omega_{total}\equiv \Omega_T$ (defined by the
ratio between the total matter and the critical energy density),
close to one. We should note that each cosmological model is
defined by its content of matter, which is reflected by the value
of $\Omega_T$. On the other hand, it is well known that this
parameter is related to the geometry of the universe: $\Omega_T >
1$ (closed), $\Omega_T = 1$ (flat) and $\Omega_T < 1$ (open). At
the moment, we do not know precisely the amount of matter present
in the universe. Therefore, still we do not know yet what the
geometry of the universe is. However, combined measurements of
Cosmic Microwave Background (CMB) temperature fluctuations and the
distribution of galaxies on large scales strongly suggest that the
universe is more likely to be flat~\cite{OsSt,Be-etal}, consistent
with the standard inflationary prediction\cite{Gu}.

On the other hand, recent measurements of a type Ia supernova (SNe
Ia )~\cite{Pe-etal,Ga-etal}, at redshift $z \sim 1$, indicate that
the expansion of the present universe is accelerated. This shows
that in the universe there exists an important matter component
that, in its  most simple description, has the characteristic of a
cosmological constant, i.e. a vacuum energy density which
contributes to a large component of negative pressure. Other
possible interpretations have been given for explaining the
corresponding astronomical data. Among them, we distinguish those
related to dark energy (quintessence) models, which are
characterized by a scalar field $\chi$ and its potential
$V(\chi)$~\cite{CaDaSt}.

One characteristic of this acceleration is that it may be quite
recent, since it has been determined for low redshift parameter,
$0.5 < z < 1$. Beyond these redshifts the behavior of the universe
should be decelerating~\cite{FiRi}. In this way, at different
epochs  in the evolution of the universe we may expect that
different "matter" components dominated its evolution. Certainly,
these different  "matter" components should be compared not only
with each other, but also with the curvature term, which defines
the geometry of the universe.

In relation to the geometry of the universe, most cosmologists
prefer a flat rather than a closed or an open universe. This is
motivated by the mentioned redshift-distant relation for supernova
of type Ia, anisotropies in the cosmic microwave background
radiation~\cite{RoHa,Me-etal} and gravitational lensing~\cite{Me},
which suggest~\cite{Be-etal} that $ \ds \Omega_T = \Omega_M +
\Omega_{\Lambda} = 1.00 \pm 0.12 \,\,\,\,(\,95\%\,\, cl\,)
\label{OT}$, with $\Omega_M$ the ordinary matter density
parameter, where baryons and CDM are the main contributions, and $
\ds \Omega_{\Lambda}\, \equiv \,\frac{\Lambda}{3\,H_0^2}$ is
associated with a smoothly distributed vacuum energy referred to
as a cosmological constant $\Lambda$, and with $H_0$ the actual
value of the Hubble parameter (from now on the subscript zero
refers to quantities evaluated at the current epoch). The
expression for $\Omega_T$ given above is the main characteristic
of the so called $\Lambda$CDM model.

Nothing can prevent us from thinking  that this flatness might be
due to a compensation among different components that enter into
the dynamical equations. In fact, our main goal in this paper is
to describe this idea in a Scalar-Tensor or in a version of the
Jordan-Brans-Dicke (JBD) theory\cite{JBD}, where the JBD scalar
field, $\phi$  has associated a scalar potential $V(\phi)$. The
salient feature of this sort of theory depends on the strength of
the dimensionless coupling "constant" $\omega$  that depends on
the JBD scalar field $\phi$ in general. Here, we will consider it
to be a constant that we will designated by $\omega_0$. At
present, observational limits from the solar system measurements
give $\omega_0 \gtrsim 3000$\cite{Wi}. We will fix the value of
this parameter to be the quantity $\omega_0 = 3000$, throughout
this paper.

We intend to start with a closed universe model composed of three
matter components. One of these components is the usual
nonrelativistic dust matter (baryonic and CDM ) $\rho_{_{M}}$, and
the other two components correspond to quintessence-type matters
which we will characterize by the scalar fields $Q$ and $\chi$.
The scalar field $Q$ will be introduced in such a way that its
dynamics will exactly cancel  the curvature together with the
scalar potential term associated with the  JBD field, so that the
resulting model will mimic a flat accelerating universe in a
genuine JBD theory. Both models become indistinguishable at a low
enough redshift parameter. The JBD mimicked accelerated universe
will be dominated by the scalar field $\chi$. It has the
characteristics of a dark energy component, and thus, its function
will be to produce the acceleration of the universe.

One of the main characteristics of the $Q$ field  is that it obeys
an equation of state given by $P_{_Q} = w_{_Q} \rho_{_Q}$. Here,
$P_{_Q}$ and $\rho_{_Q}$ represent the pressure and the energy
density, respectively. The quantity $w_{_Q}$ corresponds to the
equation of state parameter and its range will be determined later
on. Furthermore, we shall assume that this scalar field does not
interact with any other field, except with the gravitational. This
scalar field will appear in the fundamental field equations as a
fluid component.

On the other hand,  the scalar field $\chi$, contrarily to $Q$,
will interact not only with the gravitational field, but also with
the JBD scalar field $\phi$. The reason for this is that we want
to obtain the correct equations of motion in the limit $\chi
\rightarrow const.$, i.e. a universe model dominated by a
cosmological constant in a JBD theory~\cite{UeKi}. We will return
to this point in section V.

We should mention that other authors have treated the
quintessence problem in a JBD-type of theory. For example, in
ref.~ \cite{BeMa} the authors took a scalar potential for a JBD
field given by $V(\phi) = V_0 \phi^2$. here, they could describe
an accelerating universe but without a quintessence component. In
this respect, our model could be considered to be a generalization
of that model since, as we will see, we will take the scalar
potential to be of the form $V(\phi,\chi) = \phi V(\chi)$. As we
mention above, the dependence of this potential on terms of the
JBD scalar field was motivated by the fact that it gives in the
limit $\chi \longrightarrow const.$ a genuine cosmological
constant accelerated universe described in a JBD theory. In
ref.~\cite{BaPa1}  the model is still more simpler in the sense
that it does not consider a scalar potential for the JBD field
neither quintessence, but this model presents some problems: for
instance it can not describe nucleosynthesis in the radiation era.
Ref. ~\cite{BaPa2} apparently solves this problem. All of these
models start from a flat geometry. Our model, aside from depicting
a closed geometry, does not present this problem, since this model
is described in such a way that for $\omega \longrightarrow
\infty$ and $\phi \longrightarrow const.$ we achieve the results
obtained in Einstein's theory of gravity~\cite{SdC}. The price
that we must pay is the model's complexity where three scalar
fields have been introduced into this model.

The plan of the paper is as follows: In section II we write down
the field equations for a curved universe model in the
scalar-tensor theory. In section III we  introduce the constraint
equations that allow mimicking a flat universe model from a curved
universe model. Section IV describes a flat accelerated universe
model, in which we determine the main properties of the
accelerating scaler field. Here, we also determine the
deceleration parameter and the angular size as a function of time
and of redshift, respectively. In section V we study a universe
model dominated by a cosmological constant. Our conclusions are
drawn in section VI.


\section{\label{sec:level2}The JBD-type field equations }

We take the effective action  to be given by
$$ \hspace{-1.0cm}
\ds S\,=\,\frac{1}{16\pi}\,\int{d^{4}x\,\sqrt{-g}}\,\left
[\,\phi\,R\, +\,\frac{\omega_0}{\phi}\,\partial_{\mu}\phi
\,\partial^{\mu}\phi \right.
$$ \be  \ds \left. -\widetilde{V}(\phi)+16\pi\left(\frac{1}{2}\partial_{\mu}\chi
\partial^{\mu}\chi-\phi V(\chi)+ {\cal{L}}_{Fluid}\right){\frac{}{}}^{}
\right ], \label{ac1} \en where $R$ is the Ricci scalar curvature,
$\widetilde{V}(\phi)$ is the scalar potential associated with the
scalar JBD field $\phi$, and ${\cal{L}}_{Fluid}$ is a classical
bicomponent-fluid Lagrangian in which we include the minimally
coupling scalar field $Q$ and the CDM component. From now on we
disregard any possible coupling of $Q$ with ordinary matter,
radiation, or dark matter. Notice that we have included here an
interaction between the JBD field $\phi$, and the dark energy
field $\chi$. The inclusion of this interaction comes motivated by
the fact that we want to associate the scalar potential $V(\chi) $
with the cosmological constant $\Lambda$, in the limit $\chi
\longrightarrow const.$, together with the correspondence $\phi R
\longrightarrow \phi(R-2\Lambda)$. Another possibility is to
consider the potential associated with the JBD field
$\widetilde{V}(\phi) $, with a variable cosmological constant.
Here we follow the former approach in which the cosmological
acceleration is completely due to the scalar field $\chi$.

The variation of the action~(\ref{ac1}) with respect to the metric
tensor $g_{_{\mu \nu}}$, the JBD field $\phi$, and the
Quintessence-like scalar field $\chi$, yields the following set of
Equations:
$$ \ds G_{\mu \nu} = -\frac{8 \pi}{\phi}T^M_{\mu \nu} -
\frac{\omega_0}{\phi^2}\left[\partial_\mu \phi \partial_\nu \phi -
\frac{1}{2}g_{_{\mu \nu}}\partial^\alpha \phi \partial_\alpha \phi
\right]$$ \be \ds -\frac{1}{\phi}\left[D_\mu D_\nu \phi - g_{_{\mu
\nu}}\Box \phi + \frac{1}{2}g_{_{\mu \nu}}\widetilde{V}(\phi)
\right] \label{g}\en
$$ \ds - \frac{8 \pi}{\phi} \left[ \partial_\mu \chi \partial_\nu \chi
- \frac{1}{2}g_{_{\mu \nu}}\partial^\alpha \chi \partial_\alpha
\chi + g_{_{\mu \nu}}\phi V(\chi)\right],$$

$$ \hspace{-2cm} \ds \Box \phi -\frac{1}{2 \phi}\partial_\alpha \phi \partial^\alpha \phi +
\frac{\phi}{2 \omega_0}\frac{\partial \widetilde{V}(\phi)
}{\partial \phi} $$ \be \hspace{4cm} \ds = -\frac{\phi}{2
\omega_0}\left[16 \pi V(\chi) -R \right], \label{phi}\en \be \ds
\Box \chi + \phi \frac{\partial V(\chi)}{\partial \chi} = 0
\label{chi}\en and, for each fluid component we have a
conservation law \be \ds T^{i\,\,\,\,;\nu}_{\mu \nu} = 0
\hspace{1cm} (i = 1,2).\label{ce} \en In these Equations, $G_{\mu
\nu}$ is the Einstein tensor, $R$ the scalar curvature, $T^M_{\mu
\nu}$ the matter stress tensor associated with the two fluid
lagrangian, $ {\cal{L}}_{Fluid} $, and $\ds \Box \equiv D_\alpha
D^\alpha = \frac{1}{\sqrt{-g}}\frac{\partial}{\partial
x^\alpha}(\sqrt{-g}g^{\alpha \beta}\partial_\beta)$.

If we assume that the spacetime is isotropic and homogeneous with
metric corresponding to the standard Friedman-Robertson-Walker
(FRW) metric
$$ \hspace{-3.0cm} \ds d{s}^{2}= d{t}^{2}- a(t)^{2}\, \left [
\frac{dr^2}{1-k r^2} \right] $$ \be \hspace{2.0cm} \left.  +\,r^2
\left(\,d\theta^2+ sin^2 \theta \,d\varphi^2 \right)
\frac{}{}\,\right ], \label{me1} \en  where  $a(t)$ represents the
scale factor and  the parameter $k$ takes the values
$k\,=\,-\,1,\, 0,\, 1$ corresponding to an open, flat, and closed
three-geometry, respectively, and considering also that the JBD
field $\phi$ is homogeneous, i.e. is a time-depending quantity
only, (the same is assumed for the other fields) the set of
Equations~(\ref{g}) -~(\ref{chi})  yields  the following field
Equations:
$$  \ds 3\,\left(\,\frac{\dot{a}}{a}\,\right)^{2}\,
+3\,\frac{k}{a^{2}}\,=\,\frac{1}{2\phi}\, \left
[\frac{}{}\,\frac{\omega_0}{\phi}(\dot{\phi})^2\,+\,\widetilde{V}(\phi)
-6\, \frac{\dot{a}}{a}\,\dot{\phi} \,\right.$$ \be \hspace{2.0cm}
\ds \left. +16\pi\,\left( \rho_{_{Fluid}} +
\frac{1}{2}\dot{\chi}^2 + \phi V(\chi)\right )\, \right ] ,
\label{g00} \en

$$\hspace{0.0cm} \ds 2\,\frac{\ddot{a}}{a}\,
\,+\,\left(\,\frac{\dot{a}}{a}\,\right)^{2}\,+\,\frac{k}{a^{2}}\,=-\,\frac{1}{2\phi}\,
\left [\frac{}{}\,\frac{\omega_0}{\phi}(\dot{\phi})^2 +
2\,\ddot{\phi}\,+4\,\frac{\dot{a}}{a}\,\dot{\phi}\right.$$ \be
\hspace{1.0cm} \ds \left.
 -\,\widetilde{V}(\phi)\,+16\pi\,\left (p_{_{Fluid}} +
  \frac{1}{2} \dot{\chi}^2 - \phi V(\chi)\right ) \, \right ] , \label{g11}
\en
$$ \hspace{-2.0cm}\ds \frac{\ddot{\phi}}{\phi}\,+\,3\,
\frac{\dot{a}}{a}\,\frac{\dot{\phi}}{\phi}\,+\,\frac{1}{3 +
2\omega_0}\left[\frac{\partial \widetilde{V}(\phi)}{\partial \phi}
 - \frac{2 \widetilde{V}(\phi)}{\phi}\right] $$ \be =  \ds \frac{8
\pi}{3 +
2\omega_0}\,\frac{1}{\phi}\,\left[\rho_{_{Fluid}}-3p_{_{Fluid}} -
\dot{\chi}^2 + 2 \phi V(\chi) \right] ,\label{fi1}\en and \be
\ddot{\chi} + 3 \frac{\dot{a}}{a} \dot{\chi} = - \phi
\frac{\partial V(\chi)}{\partial \chi}\label{chi1}. \en

From Eq.~(\ref{ce}), the continuity equations for each individual
fluid component are given by \be \ds \dot{\rho}_{_i}\,=\,
-3\,\frac{\dot{a}}{a}\,\left(\rho_{_i}
+P_{_i}\right)\hspace{1cm}(i=1,2) \label{eqs1}.\en

These fluid components represent, on the one hand, the usual (Cold
Dark) matter component ($\rho_{_M}$,$P_{_M}$) with equation of
state $P_{_M}=0$ (dust), and therefore $\rho_{_M}(t) \propto
a^{-3}(t)$. On the other hand, we have the quintessence-like
scalar field $Q(t)$. For this latter field we define a density and
pressure by \be \ds \rho_{_{Q}} = \frac{1}{2}\dot{Q}^2\, +
\,V(Q),\label{rho1} \en and by \be \ds P_{_{Q}} =
\frac{1}{2}\dot{Q}^2\, -\, V(Q),\label{p1} \en respectively. This
scalar field obeys an equation of state given by $P_{_{Q}} =
w_{_{Q}} \rho_{_{Q}}$. In the following, we consider this
parameter to be constant, and we will determine its range of
values in the next section. Here also, the gravitational
"constant" becomes given by \be \ds G(t) = 2\left(\frac{\omega_0 +
2}{2 \omega_0 + 3}\right) \frac{1}{\phi(t)}\label{gc} .\en This
latter expression fixes the present value of the JBD scalar field
$\phi$ in terms of the Newton constant $G$ and the JBD parameter
$\omega_0$.

\section{\label{sec:level3}The flatness constraint equations}

As was specified in the introduction, we want to describe a curved
universe  which, at low redshift, mimicks a flat universe model.
In order to do this, we substitute the fluid components into the
field Equations, and we extract the following "flatness constraint
Equations": \be \ds 3\frac{k}{a^2} = \frac{8
\pi}{\phi}\,\rho_{_Q}+\frac{\widetilde{V}(\phi)}{2\,\phi},
\label{fc1}\en \be \ds -\frac{k}{a^2} = \frac{8
\pi}{\phi}\,P_{_Q}-\frac{\widetilde{V}(\phi)}{2\,\phi}
\label{fc2},\en and \be \ds \frac{8 \pi}{\phi}\,
\left(\rho_{_Q}-3P_{_Q}\right)\,=\,\frac{\partial
\widetilde{V}(\phi)}{\partial \phi}
-2\frac{\widetilde{V}(\phi)}{\phi}. \label{fc3}\en With these
conditions the following set of dynamical field equations occurs:
$$\hspace{-2.0cm} \ds 3\,\left(\,\frac{\dot{a}}{a}\,\right)^{2}\,
- 8 \pi\,V(\chi)=\,\frac{8 \pi}{\phi}\, \left [\frac{}{}\,
\rho_{_{M}}\,+\frac{1}{2}\dot{\chi}^2 \right]$$ \be \hspace{2.0cm}
\ds
+\,\frac{\omega_0}{2}\left(\frac{\dot{\phi}}{\phi}\right)^2\,-\,3\,
\frac{\dot{a}}{a}\,\frac{\dot{\phi}}{\phi}, \label{g00-1} \en

$$ \hspace{-3.5cm} \ds - 2 \frac{\ddot{a}}{a} -
\left(\,\frac{\dot{a}}{a}\,\right)^{2} + 8 \pi\,V(\chi)=\,\frac{4
\pi}{\phi}\,\dot{\chi}^2$$ \be \ds \hspace{3cm}
+\,\frac{\omega_0}{2}\left(\frac{\dot{\phi}}{\phi}\right)^2\,
+\,\frac{\ddot{\phi}}{\phi}\,+\,
2\,\frac{\dot{a}}{a}\,\frac{\dot{\phi}}{\phi} \label{g11-1}, \en
and \be \ds \frac{\ddot{\phi}}{\phi}\,+\,3\,
\frac{\dot{a}}{a}\,\frac{\dot{\phi}}{\phi}\,=\,\frac{8\,\pi
}{3+2\,\omega_0}\left
[\frac{1}{\phi}\left(\rho_{_M}-\dot{\chi}^2\right)+
2\,V(\chi)\right]\label{fi11}. \en To this set of Equations we
should add Eq.~(\ref{chi1}) which does not change. Therefore, our
fundamental set of Equations is formed by
Eqs.~(\ref{g00-1})~-~(\ref{fi11}) together with Eq.~(\ref{chi1}).
We should stress here that this set of Equations correspond to a
genuine JBD theory. Therefore, we have passed from a curvature
(closed) model, described by using a scalar-tensor theory
(characterized by a scalar potential), to a flat model, described
by using the JBD theory.

{
At this point we might specify that  the constraint
Eqs~(\ref{fc1}) -~(\ref{fc3}) may be looked upon as assumptions
meant for some simplification of the Equations of motions
described by Eqs~(\ref{g00}) -~(\ref{fi1}).

Before studying this set of Eqs. of motion we want to describe the
characteristics of either of the scalar fields $Q$ and $\phi$ by
using the constraint Eqs.~(\ref{fc1}) -~(\ref{fc3}).

First of all, adding Eqs.~(\ref{fc1}) and~(\ref{fc2}) and using
the Equation of state $P_{_{Q}} = w_{_{Q}} \rho_{_{Q}}$, we obtain
an expression that, when evaluated at present time, we get \be \ds
\Omega_Q = \frac{4}{3 \kappa}\left(
\frac{\Omega_k}{1+w_{_{Q}}}\right) ,\label{O1} \en where $\kappa$,
$\Omega_Q$ and $\Omega_k$  become defined by $\kappa
=\frac{3+2w_0}{2+w_0}$, $\Omega_Q\, =\,\left
(\frac{8\,\pi\,G}{3\,H_0^2} \right )\rho_{_Q}^0$ and $\Omega_k =
\frac{k}{H_0^2 a_0^2} $, respectively. Now, since $k \neq 0$ and
$\mid w_{_{Q}}\mid < 1$, we should take $k=1$, otherwise the
energy density $\rho_{_{Q}}$ would be negative, violating the
strong energy condition. Therefore, in the following we will
restrict ourselves to closed universe models. On the other hand,
by subtracting Eqs.~(\ref{fc1}) and~(\ref{fc2}) and using again
the Equation of state for the $Q$ field, we get \be \ds
\rho_{_{Q}} = \frac{1}{8 \pi} \left(\frac{1}{\sigma}\right)
\widetilde{V}(\phi)\label{ro1}, \en  where $\sigma$ becomes
defined by $\sigma = 1+3 w_{_{Q}}$. This latter expression tells
us that we must have $w_{_{Q}}
> -1/3$ in order to make $\rho_{_{Q}}$ a positive quantity. Thus,
we can say that the range of the $w_{_{Q}}$ parameter corresponds
to $-1/3 < w_{_{Q}} \leq 1$.

Combining Eqs.~(\ref{fc3}) and~(\ref{ro1}), we can obtain an
explicit expression for the JBD scalar potential: \be \ds
\widetilde{V}(\phi) = \widetilde{V}_0 \left
(\frac{\phi}{\phi_0}\right )^{3\left (
\frac{1+w_{_{Q}}}{\sigma}\right)}. \label{po1} \en Here, the
constant of integration has been fixed by asking that the scalar
potential at present time becomes $\widetilde{V}_0$.

With the help of this latter relation we could get a more precise
value of the constant $w_{_{Q}}$. By using Eq.~(\ref{po1})
together with the constraint Eqs.~(\ref{fc1}) and~(\ref{fc2}) we
get a relation between the scale factor $a$ and the JBD field
$\phi$ given by $ \ds \left(\frac{1+w_{_{Q}}}{\sigma} \right) a^2
=
\frac{\phi_0}{\widetilde{V}_0}\left(\frac{\phi}{\phi_0}\right)^{-2/\sigma}$,
which yields the following relation: \be \ds H(t) =
\left(\frac{1}{\sigma}\right)\,\epsilon(t) ,\label{he} \en where
$H(t)=\dot{a}/a$ is the hubble parameter and $\epsilon(t)=
-\dot{\phi}/\phi=\dot{G(t)}/G(t)$ represents the changing rate of
the gravitational constant. When this expression is evaluated at
present time, we get for the parameter $w_{_{Q}}$ the following
quantity: \be \ds w_{_{Q}} = - \frac{1}{3} +\frac{1}{3}
\frac{\epsilon_{_{0}}}{H_0}.\label{q} \en Therefore, this
parameter becomes determined by the ratio of the present time
variation of the gravitational constant and the Hubble parameter.
Local laboratory and solar system experiments  put an upper limit
on the $\epsilon_{_{0}}$ parameter given by $\epsilon_{_{0}} \leq
\pm 10^{-11}$ per year~\cite{Gi}. This limit together with the
value measured for the Hubble parameter, which is in the range
$56$~[Km/s/Mpc]~$\leq H_0 \leq 88$~[Km/s/Mpc]  according to the $2
\sigma$ range of the HST Key Project~\cite{Fretal}, induce a limit
for the $w_{_{Q}}$ parameter. We will consider a positive value
for $\epsilon_{_{0}}$ (the gravitational "constant" is an
increasing function of time). In order to agree with the bound
specified for $w_{_{Q}}$ previously, we will use the value
$w_{_{Q}}=-0.3324$, which corresponds to $\epsilon_{_{0}} =
10^{-13}$ per year and $H_0 = 72$~[Km/s/Mpc].

Since our interest is to describe an accelerating universe , we
will assume that $a$ is a function of the cosmic time $t$ in the
form $ a(t) = a_0 \left(\frac{t}{t_0}\right)^N$, with $N \gtrsim
1$. With this assumption, together with the constraint Eqs. and
the Equation of state for the scalar field $Q$, we find that $\ds
Q(t) = \overline{Q}\left[1-
\left(\frac{t_0}{t}\right)^{\alpha-1}\right] + Q_0$ and $\ds
V_Q(t) = V_Q^0 \left(\frac{t_0}{t}\right)^{2\alpha}$, where $Q_0$
is the present value of $Q$,  $ \ds \alpha =
\frac{3}{2}N(1+w_{_{Q}})$, $\ds  \overline{Q}= \left(\frac{\alpha
- 1}{t_0}\right)\sqrt{\frac{8 \pi}{\widetilde{V}_0} \left(
\frac{\sigma}{1 + w_{_{Q}}}\right)}$, and $\ds  V_Q^0 =
\frac{\widetilde{V}_0}{16 \pi} \left(\frac{1 -
w_{_{Q}}}{\sigma}\right)$. These relations allow us to write an
explicit expression for the scalar potential associated with $Q$:
\be \ds V(Q) = V_Q^0 \left[1 -  \overline{Q}\left( Q -
Q_0\right)\right]^{\frac{2 \alpha}{\alpha - 1}} .\label{vq} \en
Figure~(\ref{Qpoten}) shows the form of the potential for four
different values of N. We note that the potential decreases when
$Q$ increases, tending to a vanishing value for $Q \longrightarrow
\infty$. Note also that as we increase the value of N, $V(Q)$
tends to zero faster. Asymptotically, this scalar field behaves as
a stiff fluid ($P_Q = \rho_{_{Q}}$), for $\dot{Q} \neq 0$.
\begin{figure}[ht]
\includegraphics[width=3.0in,angle=0,clip=true]{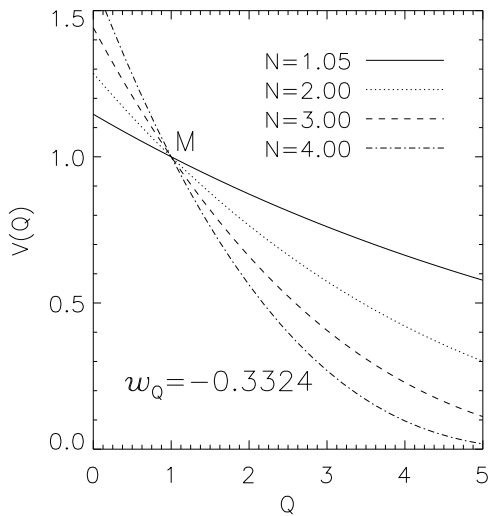}
\caption{ Plot of the scalar potential $V_Q$ (in units of $
 \frac{\widetilde{V}_0}{16 \pi}\left(\frac{1 - w_{_{Q}}}{\sigma}\right)$)
  as a function of the scalar field
$Q$ (in unit of $ Q_0 $), for four different values of the
parameter $N$. We have taken $w_{_{Q}}=-0.3324$. The point $M$,
where all the curves intercept, coincides with $\ds Q_0 \equiv t_0
\sqrt{\widetilde{V}_0/8 \pi}$.}\label{Qpoten}
\end{figure}

\section{\label{sec:level4}An accelerated universe model}

Starting from the field Eqs.~(\ref{g00-1})~-~(\ref{fi11}),
together with Eq.~(\ref{chi1}), we want to describe the main
characteristics of our accelerated model. Specifically, we want to
determine the characteristics of the scalar field $\chi$. By using
the accelerated power law solution, it is not hard to find that
the scalar field $\chi$ and its associated potential $ V(\chi)$
become given by $$ \hspace{-4cm} \chi(x)  = \sqrt{\frac {1}{\pi
\,G}}\,\, \frac{\eta_{_{1}}}{{2 \beta} }\,x^{ - \beta}$$ \be \ds
\times\,_2F_1\left([ - { \frac {\beta}{\gamma }} \,{- \frac
{1}{2}} ], \,[{ \frac {\gamma - \beta}{\gamma }} ], \,A_1\,x
^{\gamma }\right)\label{chi2}\en and \be \ds
V(x)=\left(\frac{H_0^2}{16 \pi}\right) \eta_{_{2}}\left[
x^{-2/N}-A_1 x^{-\delta}\right],\label{vx} \en respectively. Here,
the constants are given by $\beta=3(\frac{1}{2}+w_{_{Q}})$,
 $\gamma=-2(1-1/N)+3w_{_{Q}}$,  $\delta=2-3w_{_{Q}}$, $\ds \eta_{_{1}}=\sqrt{\eta_{_{2}}\,\kappa\,
(1+w_{_{Q}})}$, $\ds \eta_{_{2}}=3-\frac{w_0}{2}\sigma^2 -
3\sigma$ and $\ds A_1=\frac{3}{2}\kappa
\frac{\Omega_M}{\eta_{_{2}}}$. The minimal value of the scalar
field $\chi$, where the potential $V(\chi)$ vanishes, corresponds
to the value $$\ds \chi_{_{Min}}=\left\{\frac{3
\kappa}{2}\left(\frac{\Omega_M}{3-\frac{w_0}{2}\sigma^2 -3\sigma}
\right)\right\}^{1/(3-\sigma -2/N)}.$$
\begin{figure}[ht]
\includegraphics[width=3.0in,angle=0,clip=true]{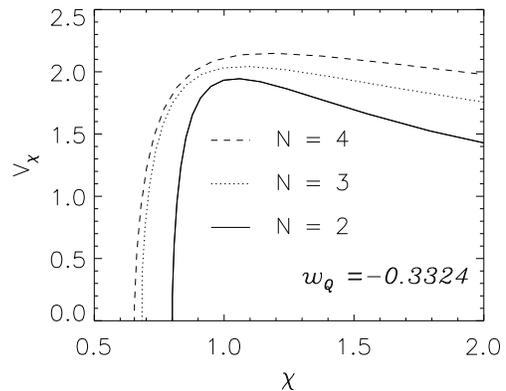}
\caption{ This graph shows the scalar potential $V_{_{\chi}}$ (in
units of $(16 \pi t_0^2)^{-1}$ ) as a function of the scalar field
$\chi$ (in units of $(2 \pi G_N)^{-1/2}$) for different values of
the parameter $N$, as is shown in the figure. Here, we have taken
$\Omega_M=0.35$, $w_{_{Q}} =-3324$ and $w_0 =
3000$.}\label{V(chi)}
\end{figure}

By using numerical computations we can plot the scalar potential
$V_Q\chi$ as a function of the scalar field $Q\chi$.
Figure~(\ref{V(chi)}) shows the potential $V(\chi)$ as function of
the scalar field $\chi$ for four different values of the parameter
$N$, and the other parameters $w_0$ and $w_{_{Q}}$ have been fixed
at 3000 and -0.3324, respectively. All of these curves
asymptotically tend to vanish for $\chi\longrightarrow \infty $.
There, in this limit and for $\dot{\chi} \neq 0$, the universe
becomes dominated by a stiff fluid with Equation of state $P_\chi
= \rho_{_{\chi}}$. This property also is found in similar models
described by using Einstein's theory of gravity~\cite{SdC}.

If we associate with the scalar field $\chi$ an energy   and
pressure density defined  by $\rho_{_{\chi}}=
\frac{1}{2}\dot{\chi}^2 + \phi V(\chi)$ and $P_\chi=
\frac{1}{2}\dot{\chi}^2 - \phi V(\chi)$, respectively, we can
introduce an equation of state parameter $w_{_{\chi}}$ defined by
the ratio $P_\chi/\rho_{_{\chi}}.$ It is not hard to see that this
quantity is negative for small redshifts. This means that the
scalar field $\chi$ acts as the source for the acceleration of the
universe. In general this quantity is variable and becomes a
constant only for $N=\frac{2}{2-3w_{_{Q}}}$. In any case its
present value will be interesting at the time of calculating the
present deceleration parameter $q_0$, as we will see soon.

From the definition of the deceleration parameter $\ds
q(t)=-\frac{\ddot{a}(t)}{a(t) H(t)^2}$ together with the field
Equation of motion, Eqs.~(\ref{g11-1}), we obtain   $$
\hspace{-5cm} \ds q(t)=\frac{1}{2} + \frac{4 \pi}{3 + 2
w_0}\frac{\rho_{_{M}}}{H^2 \phi} $$ \be \ds +4 \pi
w_{_{\chi}}\left(\frac{1+2w_0}{3+2w_0}\right)
\frac{\rho_{_{\chi}}}{H^2 \phi} +
\frac{w_0}{4}\left(\frac{\epsilon}{H}\right)^2 +
\frac{1}{2}\left(\frac{\epsilon}{H}\right).\label{qt} \en This
expression, together with Eq.~(\ref{g00-1}), when evaluated at
present time, gives
$$\ds q_{_{0}} = \frac{1}{2}\left(
\frac{3+w_0}{2+w_0}\right)\Omega_M + \frac{1}{4} \kappa
\Omega_\chi\left[\frac{}{} 1\right.$$ \be \ds \left.+ 3\left(
\frac{1+2w_0}{3+2w_0}\right) w_{_{\chi}}^0\right
]+\frac{w_0}{3}\left(\frac{\epsilon}{H}\right)_0^2 +
\left(\frac{\epsilon}{H}\right)_0,\label{q0} \en where
$w_{_{\chi}}^0$ represents the present equation of state
parameter. Note that we have introduced the present density
parameter $\Omega_\chi $ defined by $\Omega_\chi = \frac{8 \pi
G}{3 H_0^2}\rho_{_{\chi}}^0$. In order to describe an acceleration
for our model, we need to satisfy for the $w_{_{\chi}}^0$
parameter the following inequality: $$ \ds w_{_{\chi}}^0 < -
\frac{1}{3}\left( \frac{3+2w_0}{1+2w_0}\right)\left[  1 + 2\left(
\frac{3 +
w_0}{3+2w_0}\right)\frac{\Omega_M}{\Omega_\chi}\right],$$ which
coincides with the result found in Einstein´s theory of gravity,
in the limit $\omega_0 \longrightarrow \infty $. Note that here we
have neglected the ratio $\ds \left( \frac{\epsilon}{H}
\right)_0$, (and its square).

Now we would like to calculate the angular size  $\Theta(z)$ as a
function of the redshift $z$. In order to do this, we need first
to calculate the luminosity distance $d_L(z)$. This parameter
plays a crucial role in describing the geometry and matter content
of the universe. From the metric~(\ref{me1}), we observe that
light emitted by the object of luminosity ${\cal{L}}$ and located
at the coordinate distance $\theta$ at a time $t$, is received by
an observer (assumed located at $\theta = 0$) at the time $t=t_0$.
The time coordinates are related by the cosmological redshift $z$
in the $\theta$ direction by the expression: $1+z = a_0/a(t)$. The
luminosity flux reaching the observer is ${\cal{F}} = {\cal{L}}/4
\pi d^2_L$, where $d_L$ is the luminosity distance to the object,
given by $d_L(z)=a_0 sin[\theta(z)] (1+z)$. On the other hand, if
we want to obtain an explicit expression for the angular size, let
us now consider an object aligned to the $\varphi$ direction and
proper length $l$, so that its "up" and "down" coordinates are
$(\theta, \varphi +\delta \varphi, 0)$ and $(\theta, \varphi, 0)$.
The proper length of the object is obtained by setting $t =$
const. in the line-element metric~(\ref{me1}), $ds^2 = -l^2 =
-a^2(t) sin^2(\theta)\delta \varphi ^2$. Thus, the angular size
becomes $\ds \delta \varphi\equiv \Theta(z)=
\frac{l}{d_L(z)}\,(1+z)$. Our solution gives \be \ds \Theta(z) = l
H_0 \frac{ \overline{\alpha} (1+z)}{\sin\left\{\overline{\alpha}
\left[ (1+z)^{\overline{N}} -1 \right] \right\}},\label{as}\en
where $\overline{N} = 1 -1/N$ and $\overline{\alpha}= \frac{3}{4}
(1+w_{_{Q}})\kappa \Omega_Q$.

Figure~\ref{ang} shows the angular size as a function of the
redshift for three different values of the parameters N (N=2, 3,
4). We have used the value $\Omega_Q = 0.02$. For comparison, in
this plot we have added the graph of the angular size
corresponding to the flat FRW model for N=4.  Notice that, at
sufficient large redshift, the two curves begin to separate.
Therefore, we expect that we could distinguish them at high enough
redshift.

\begin{figure}[ht]
 \includegraphics[width=3.0in,angle=0,clip=true]{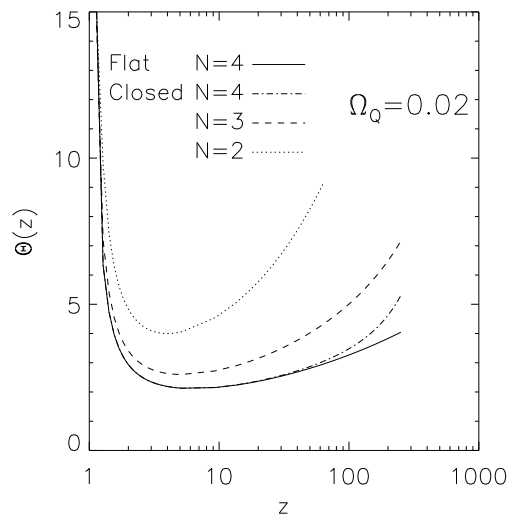}
\caption{ This plot shows the angular size $\Theta(z)$ (in units
of $l H_0$) as a function of the redshift $z$, for different
values of the parameter $N$ (N=2,3 and 4). We have included in
this plot the angular size for a flat model in the case $N=4$.
Notice that at low enough redshift the models become
indistinguishable.} \label{ang}
 \end{figure}

\section{\label{sec:level5}cosmological constant universe model}

Here we study the case in which  $\chi =\chi_0= const.$ and taking
$ 8 \pi V(\chi_0)=\lambda$. The set of Equations~(\ref{g00-1})
-~(\ref{fi11}) reduces to
\be \ds 3\,\left(\,\frac{\dot{a}}{a}\,\right)^{2}\, -
\,\lambda\,=\,\frac{8 \pi}{\phi}\,\rho_{_{M}}\,
+\,\frac{\omega_0}{2}\left(\frac{\dot{\phi}}{\phi}\right)^2\,-\,3\,
\frac{\dot{a}}{a}\,\frac{\dot{\phi}}{\phi} \label{g00-11} \en

\be \ds - 2 \frac{\ddot{a}}{a} -
\left(\,\frac{\dot{a}}{a}\,\right)^{2}\,+\,\lambda
 =\frac{\omega_0}{2}\left(\frac{\dot{\phi}}{\phi}\right)^2\,
+\,\frac{\ddot{\phi}}{\phi}\,+\,
2\,\frac{\dot{a}}{a}\,\frac{\dot{\phi}}{\phi} \label{g11-11}, \en
and \be \ds \frac{\ddot{\phi}}{\phi}\,+\,3\, \left
(\frac{\dot{a}}{a}\right)\,\left(
\frac{\dot{\phi}}{\phi}\right)\,=\,\frac{2 \lambda
}{3+2\,\omega_0} + \frac{8 \pi}{\phi}
\frac{\rho_{_M}}{3+2\,\omega_0} \label{fi111}. \en This set of
Eqs. coincides with that studied in Ref.~\cite{UeKi}. There, dust
($P_M \approx 0$) was considered for the matter component. The
case $\Lambda =0$ was treated in Ref~\cite{NCSdCRH}.

In order to write down a possible solution for the set of
Equations~(\ref{g00-11}) -~(\ref{fi111}) we first notice that the
quantity
$$\ds B_1=\sqrt{1-\frac{2}{4+3w_0}\left(\frac{2}{\kappa}\right)
\left(\frac{\Omega_\chi}{\Omega_M^2}\right)\left[1+(1+w_0)\sigma\right]^2}
,$$ with $\Omega_\chi$, $\Omega_M$, $\kappa$ and $\sigma$ defined
in the previous section, is real for the present values of the
quantities that we are considering. Therefore, the solution to
this set of Equations which vanishes at $t=0$ becomes~\cite{UeKi}

\begin{eqnarray}
 \nonumber a(t)&=&a_0\,\left\{\frac{B_1 \cosh\left[2
\kappa_{_{\chi}}(x-x_c)\right]-1} {B_1 \cosh\left[2
\kappa_{_{\chi}}(x_0-x_c)\right]-1}\right\}^{\alpha_1} \\
\nonumber && \\
 & & \hspace{1.cm} \ds
 \times \left\{\, \ds \frac{\frac{B_2\, \tanh\left[\,
\kappa_{_{\chi}}\,(x\,-\,x_c)\,\right]\,-\,1}{B_2\, \tanh\left[\,
\kappa_{_{\chi}}\,(x\,-\,x_c)\,\right]\,+\,1}}{\frac{B_2\,
\tanh\left[\,
\kappa_{_{\chi}}\,(x_0\,-\,x_c)\,\right]\,-\,1}{B_2\,
\tanh\left[\, \kappa_{_{\chi}}\,(x_0\,-\,x_c)\,\right]\,+\,1}}
\right\}^{\alpha_2}, \label{al}
\end{eqnarray}
where $x=H_0\,t$, $\ds \kappa_{_{\chi}} =
\frac{1}{2}\overline{\eta} \sqrt{3\Omega_\chi}$, $\ds B_2
=\sqrt{\frac{1+B_1}{1-B_1}}$, $\ds x_c =
-\frac{1}{\kappa_{_{\chi}}}\tanh^{-1}(B_2^{-1}) $, $\ds
\alpha_1=\frac{1+w_0}{4+3w_0}$, $\ds
\alpha_2=\frac{1}{4+3w_0}\sqrt{\frac{3+2w_0}{2}}$ and $\ds
\overline{\eta}=\sqrt{2\left(\frac{4+3w_0}{3+2w_0}\right)}$.

From this expression we can determine the deceleration parameter
$q(t)$ as we defined it previously. The following Figure shows how
this parameter changes with time for three different values of the
density parameter $\Omega_\chi = \frac{16 \pi V_\chi^0}{3 H_0^2}$.
\begin{figure}[ht]
 \includegraphics[width=3.0in,angle=0,clip=true]{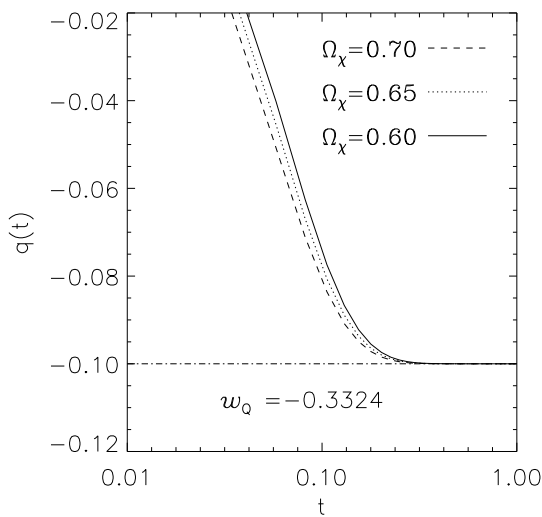}
\caption{ This plot shows the deceleration parameter $q(t)$  as a
function of  time $t$ (in unit of $H_0$), for three different
values of the parameter $\Omega_\chi$ (0.60, 0.65, 0.70).}
\label{qtjbd}
 \end{figure}

We should note that, when a value for the parameter $\Omega_\chi$
is given, the parameter $\Omega_M$ becomes immediately determined,
since this parameter has to satisfy the relation
$$\ds \Omega_M = \frac{2}{\kappa}\left( 1 -
\sigma\left[1+\frac{w_0}{6}\sigma\right]\right) - \Omega_\chi.
$$
This relation results from the field Eq.~(\ref{g00-11}) when it is
evaluated at the present time. Note also that this relation
reduces to the expression $\Omega_M + \Omega_\chi = 1$, in the
limit of Einstein's theory.

Note also that the deceleration parameter tends to the value -0.1
at the current epoch, as shown from Fig.~\ref{qtjbd},implying that
the expansion of the universe is accelerating rather than slowing
down. This certainly agrees with evidence coming from type Ia
supernova observational data~\cite{Pe-etal,Ga-etal}.
Finally, we should mention that, in the limit of $\phi
\longrightarrow const.$ and $\omega \longrightarrow \infty$, the
solution of Eq.~\ref{al} becomes \be \ds a(t)\sim
sinh^{2/3}\left(\frac{3}{2}\sqrt{\Omega_\chi}\,H_0\,t\right)
\label{a2},\en which is nothing but the exact solution to the
Friedmann Equations~\cite{UeKi}.

\section{conclusions}

We have described a closed universe model in which, apart from the
usual Cold Dark Matter component, we have included a
quintessence-like scalar field $Q$ in a scalar-tensor theory
characterized by a scalar JBD field $\phi$ and its scalar
potential $V(\phi)$. We have fine-tuned the $Q$ component,
together with the curvature term and the scalar potential
$V(\phi)$, for mimicking a flat universe model. The resulting
model corresponds to the Quintessence (or Dark Energy) Cold Dark
Matter ($\chi CDM$) scenario described in a JBD theory of gravity,
in which the JBD parameter $w_0$ has the value $w_0 = 3000$, in
agreement with  solar experiments. The flatness conditions allow
us to obtain the properties of the scalar field $Q$ together with
the JBD scalar field $\phi$. Especially, we have determined an
explicit form for the potentials $V(Q)$ and $V(\phi)$. There, we
have imposed an effective equation of state, $P_Q =w_Q
\rho_{_{Q}}$, for the field $Q$, in which the equation of state
parameter $w_Q$ has taken the value $ w_{_{Q}} = -0.3324$. The
main characteristic of $V(Q)$ is that it decreases when $Q$
increases, going to zero asymptotically. Contrarily, the potential
$V(\phi)$ occurred as an increasing quantity when $\phi$
increases.

After applying the flatness constraint Equations for our model, we
determined the angular size, apart from the deceleration
parameter. There it was shown that, at low enough redshifts, the
curvature and the flat models become indistinguishable and, on the
other hand, that the model presents an acceleration  rather than a
deceleration. The same parameter of deceleration was determined in
the limit in which the scalar field $\chi=$ const $= \chi_0$, in
which the potential term contributed as a cosmological constant,
in the identification $16 \pi V(\chi_0) = \lambda$. The resulting
Equations of motion coincided with those studied by Uehara and
Kim. With the values used for the present quantities, we have
found that our model perfectly accommodates an accelerating
universe model in a genuine JBD theory.

\begin{acknowledgments}

The authors wish to thank F. M\"{u}ller-Hoissen for the
hospitality in G\"{o}ttingen. SdC thank MECESUP FSM 9901 grant for
financial support. PS wishes to thank Deutscher Akademischer
Austauschdienst (DAAD) for financial support. The support from
grants FONDECYT Projects N$^0$ 1030469 (SdC) and 1010485 (SdC and
PS), and by grants PUCV-DGIP N$^0$ 123.764/2003, UdeC/DI N$^0$
202.011.031-1.0 are also acknowledged. The authors are grateful to
Universidad de Concepci\'on for partial support of the 2$^{nd}$
Dichato Cosmological Meeting, where this work was started.
\end{acknowledgments}

\end{document}